\documentclass[conference]{IEEEtran}
\IEEEoverridecommandlockouts
\usepackage{cite}
\usepackage{amsmath,amssymb,amsfonts}
\usepackage{algorithmic}
\usepackage{graphicx}
\usepackage{textcomp}
\usepackage{multirow}
\usepackage{multicol}
\usepackage{xcolor}
\usepackage{soul}
\usepackage{hyperref}
\def\BibTeX{{\rm B\kern-.05em{\sc i\kern-.025em b}\kern-.08em
    T\kern-.1667em\lower.7ex\hbox{E}\kern-.125emX}}
\begin{document}


\title{Multimodality Multi-Lead ECG Arrhythmia Classification  using Self-Supervised Learning}

\author{\IEEEauthorblockN{1\textsuperscript{st} Thinh Phan}
\IEEEauthorblockA{\textit{Department of AI Convergence} \\
\textit{Chonnam National University}\\
Gwangju, South Korea 61186\\
phantrandacthinh2382@gmail.com}
\and
\IEEEauthorblockN{2\textsuperscript{nd} Duc Le}
\IEEEauthorblockA{\textit{Department of CSCE} \\
\textit{University of Arkansas,}\\
 Fayetteville, Arkansas 72703  \\
minhducl@uark.edu}
\and
\IEEEauthorblockN{3\textsuperscript{rd} Patel Brijesh}
\IEEEauthorblockA{\textit{Department of Cardiology} \\
\textit{West Virginia University}\\
Morgantown,  WV 26506 \\
brijesh.patel@wvumedicine.org}
\and
\IEEEauthorblockN{4\textsuperscript{th} Donald Adjeroh}
\IEEEauthorblockA{\textit{Department of CSEE} \\
\textit{West Virginia University}\\
Morgantown,  WV 26506 \\
don@csee.wvu.edu}
\and
\IEEEauthorblockN{5\textsuperscript{th} Jingxian Wu}
\IEEEauthorblockA{\textit{Department of ELEG} \\
\textit{University of Arkansas,}\\
 Fayetteville, Arkansas 72703 \\
wuj@uark.edu}
\and
\IEEEauthorblockN{6\textsuperscript{th} Morten Olgaard Jensen}
\IEEEauthorblockA{\textit{Department of Biomedical Engineering} \\
\textit{University of Arkansas,}\\
 Fayetteville, Arkansas 72703 \\
mojensen@uark.edu}
\and
\IEEEauthorblockN{7\textsuperscript{th} Ngan Le}
\IEEEauthorblockA{\textit{Department of CSCE} \\
\textit{University of Arkansas,}\\
 Fayetteville, Arkansas 72703  \\
thile@uark.edu}}

\maketitle
\begin{abstract}
Electrocardiogram (ECG) signal is one of the most effective sources of information mainly employed for the diagnosis and prediction of cardiovascular diseases (CVDs) connected with the abnormalities in heart rhythm. 
Clearly, single modality ECG (i.e. time series) cannot convey its complete characteristics, thus, exploiting both time and time-frequency modalities in the form of time-series data and spectrogram is needed. 
Leveraging the cutting-edge self-supervised learning (SSL) technique on unlabeled data, we propose SSL-based multimodality ECG classification. Our proposed network follows SSL learning paradigm and consists of two modules corresponding to pre-stream task, and down-stream task, respectively. In the SSL-pre-stream task, we utilize self-knowledge distillation (KD) techniques with no labeled data, on various transformations and in both time and frequency domains. In the down-stream task, which is trained on labeled data, we propose a gate fusion mechanism to fuse information from multimodality.
To evaluate the effectiveness of our approach, ten-fold cross validation on the 12-lead PhysioNet 2020 dataset has been conducted. 
\url{https://github.com/UARK-AICV/ECG_SSL_12Lead}.
\end{abstract}


\section{Introduction}
CVDs are leading causes of deaths globally. The mortality rate can be considerably reduced by early treatment if occult signals linked with CVD are detected by ECG. The ECG signals, which record cardiac electrical activities, are widely adopted to diagnose abnormal heart rhythms and intra-cardiac conduction abnormalities. 

Traditionally, cardiac feature extraction \cite{chouhan2008threshold} and pattern classifiers \cite{zabihi2017detection} are separated. Notwithstanding the decent performance, they are not applicable to real-life cases because of high time consumption and complexity. Recent DNNs-based approaches have obtained amazing research progress in various domains \cite{yamazaki2022spiking, le2021deep, zhou2021deep}. 
DNN-based ECG classification in general can be categorized into single lead classification \cite{rajpurkar2017cardiologist, le2021multi} or multiple lead classification \cite{ribeiro2020automatic, baloglu2019classification}. \emph{Our proposed method belongs to the second category}. In this group, \cite{ribeiro2020automatic} 
proposed a simple residual neural network used for classifying 6 types of abnormalities on the in-house 12-lead database, where some of its testing metrics are better than those of expert cardiologists. \cite{baloglu2019classification} trained a CNNs-based structure on multi-lead ECG data to diagnose  myocardial infraction. Due to high complexity in higher-dimensional data of 12-lead ECG, different architectures have been adopted to model time correlation among ECG sample points \cite{zhang2016colorful}, \cite{wang2019global}. Besides time-series data, time-frequency has played an important role in ECG analysis. \cite{huang2019ecg} used STFT-based spectrogram and 2D CNNs for ECG arrhythmia classification. \cite{le2021multi} proposed using STFT and stationary wavelet transform (SWT) transformations to obtain two-dimensional (2-D) matrix input suitable for deep CNNs. \cite{yildirim2018novel} proposed a novel wavelet sequence based on deep bidirectional LSTM network model. 

Furthermore, with the growing demand for medical examination and treatment, the healthcare industry steadily accumulates innumerable amounts of data but these unlabelled data might not be serviceable for most tasks. To address this limitation, we leverage the recent advanced SSL techniques, i.e., contrastive learning \cite{caron2021emerging}. 
A common workflow to apply SSL is to train the network in an unsupervised manner by learning with a pre-stream task, and then finetuning the pre-trained network on a target downstream task. 
The suitable pre-stream tasks can be considered in four categories: context-based \cite{caron2018deep}, generation-based \cite{srivastava2015unsupervised}, free semantic label-based \cite{pathak2017learning, tran2022ss}, and cross-modal-based \cite{sayed2018cross, vu2021self}. Recently, SSL-based DNNs have also been applied in ECG classification. \cite{kiyasseh2020clocs} proposed a self-designed loss to bring the representations from the same patient closer and study the shared context of individual recordings through time and scenarios. \cite{lan2021intra} customized an SSL model that understands the differences among segments from individual patients and dissimilarities between patients' recordings from same category. \cite{mehari2022self} conducted extensive experiments on four SSL methods with multiple combinations of transformations and demonstrated the improvement in macro AUC when using SSL. However, the existing SSL-based ECG classification approaches only focus on time domain with time-series data and they have not exploited the ECG characteristics in frequency domain.

\begin{figure}[!t]	
\centering
\includegraphics[width=\linewidth]{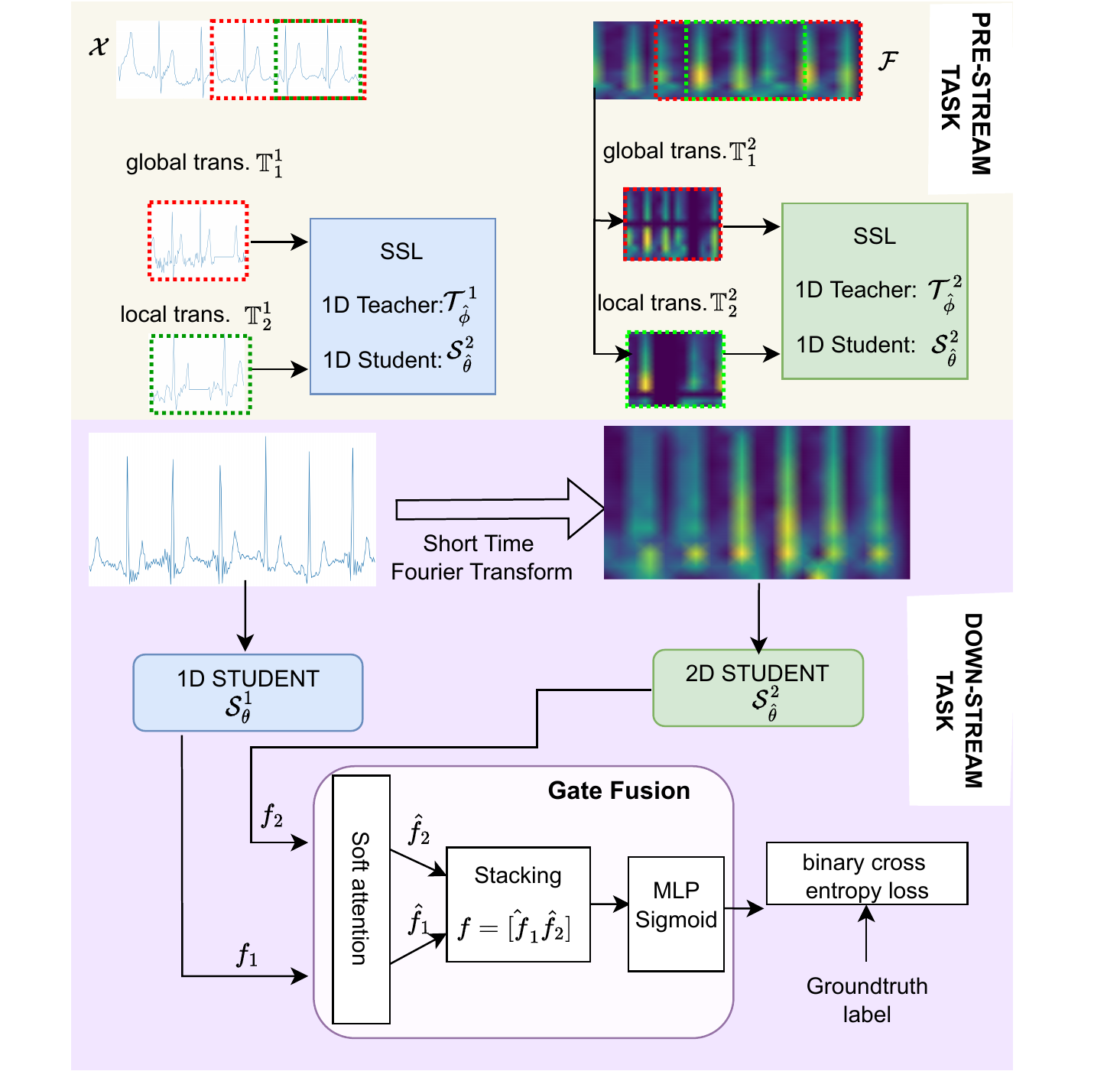}
\caption{Overall flowchart of our proposed network consisting two modules i.e., SSL pre-stream task and SSL down-stream task. In the SSL pre-stream task, the pre-trained models are student networks and they are trained on unlabeled data with self-KD mechanism. In the SSL pre-stream task, the pre-trained models are fine-tuned on labeled data. The features $f_1$ from time-series and $f_2$ from spectrogram are fused by our proposed gate fusion.}
\label{fig:flowchart}
\end{figure}

In this work, we adopt the advanced SSL technique, i.e., self-distillation SSL with no labels \cite{caron2021emerging, vu2021teaching}.
Our proposed network contains two modules i.e., SSL pre-stream task, SSL down-stream task. The first module is trained on unlabeled data and contains two components corresponding to 1D CNNs for time-series data and 2D CNNs for time-frequency spectral (spectrogram) signal. The second module fine-tunes the pre-trained models i.e. 1D CNNs and 2D CNNs from the first module to perform muti-lead ECG classification on labeled data. In the down-stream task, the features from the two networks are fused by our proposed gate fusion mechanism. The overall flowchart of our proposed network is shown in Fig. \ref{fig:flowchart}.

\vspace{-2mm}

\section{Proposed Methods}


\subsection{Self-Knowledge Distillation (KD): a revisit}
\label{subsec:self_kd}
KD \cite{hinton2015distilling} is a learning paradigm where a student network $\mathcal{S}_\theta$ is trained to match the output of a given teacher network $\mathcal{T}_\phi$, parameterized by $\theta$ and $\phi$, respectively. Given an input signal $x$, self-KD is explained in the following steps: (i) Apply different transformations ($\mathbb{T}_1$, $\mathbb{T}_2$) on  $x$ to generate global view $x_1 = \mathbb{T}_1(x)$ and local view $x_2 = \mathbb{T}_2(x)$ at different distorted views, or crops. On the same transformation, the global transformation $\mathbb{T}^1_1$ and local transformation  $\mathbb{T}^1_2$ are defined with different crop ratios i.e. greater than 50\% is considered as global; otherwise, it is local. ; (ii) Pass $x_1$ and $x_2$ to networks of $\mathcal{S}_\theta$ and $\mathcal{T}_\phi$; (iii) Compute similarity between probability distributions of output from $\mathcal{S}_\theta$  and $\mathcal{T}_\phi$ using a cross-entropy loss. Notably, the output of $\mathcal{T}_\phi$ is centered with a mean computed over the batch.

\begin{figure}[!h]	
\centering
\vspace{-2mm}
\includegraphics[width=0.4\textwidth]{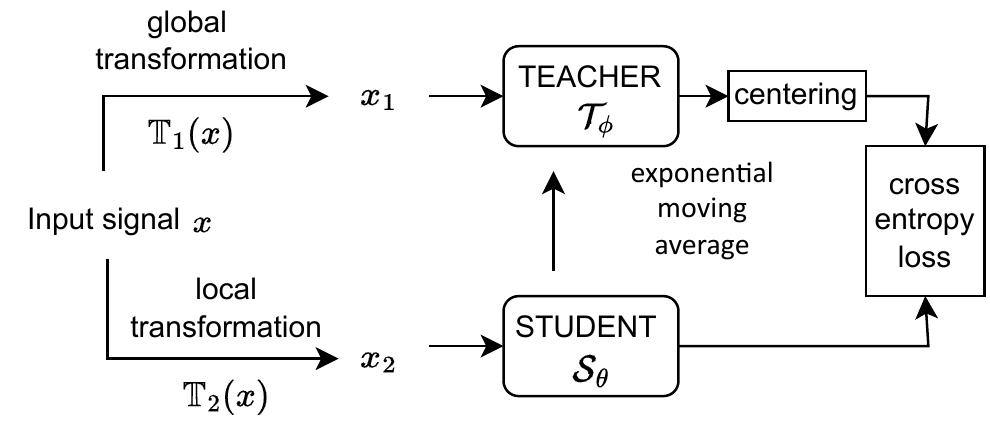}
\caption{\vspace{-2mm} Illustration of SSL with self-KD mechanism with no labeled data. \vspace{-2mm}}
\label{fig:self_kd}
\end{figure}

Unlike typical KD, Self-KD\cite{caron2021emerging} build $\mathcal{T}_\phi$ using the past iterations of $\mathcal{S}_\theta$. Thus, both $\mathcal{T}_\phi$ and $\mathcal{S}_\theta$ are sharing the same network architecture with different sets of parameters $\theta$ and $\phi$. Notably, both $\mathbb{T}_1$ and $\mathbb{T}_2$ are sharing the same transformations with different crop ratios. The overall flowchart of self-KD is illustrated in Fig. \ref{fig:self_kd}.

\subsection{SSL: Pre-stream Task}
\label{subsec:pre}
The pre-stream task makes use of self-KD technique to pre-train models of time-series data and time-frequency signal using  
two components as shown in the yellow block in Fig.\ref{fig:flowchart}. To apply self-KD technique to two different modalities, we need to define the network architectures of $\mathcal{T}_\phi$, $\mathcal{S}_\theta$ and transformations of $\mathbb{T}_1$ $\mathbb{T}_2$ as follows:

\noindent
\textbf{1. Pre-stream task with time-series data}
This component takes time-series $\mathcal{X}$ as an input. The SSL KD in this task is performed by Time Cutout (TC) and Gaussian Noise (GN) transformation $\mathbb{T}^{1}$ as follows:


Let $t_1$ and $t_2$ denote the start and the end of a random crop on signal $\mathcal{X}$, TC transformation is defined as follows: 
\vspace{-2mm}
\begin{equation}
\mathcal{X}_{i,TC} : \{X_{i}\}_{i={t}_{1}}^{{t}_{2}} = 0
\end{equation}
where $t_1 = rand(0, n -\alpha), t_2 = t_1 + \alpha$, $\alpha = \mathcal{U}(0,0.5).n$, with $n$ denoting the 
length of $\mathcal{X}$, and $\mathcal{U}$ denoting the uniform distribution. 

Let $\mathcal{N}(0,\sigma)$ denote the Gaussian Noise (GN). The 
GN transformation is defined as: $\mathcal{X}_{i,GN} : \mathcal{X}_{i} + \mathcal{N}(0,\sigma)$
The $\mathbb{T}^{1}$ transformation is $\mathbb{T}^{1} : \mathcal{X}_{i,TC} + \mathcal{X}_{i,GN}$.



We utilize xresnet1d50 \cite{he2019bag} as our backbone network for both $\mathcal{T}^1_\phi$ and $\mathcal{S}^1_\theta$, which are trained on different parameters of $\phi$ and $\theta$.

\noindent
\textbf{2. Pre-stream task with time-frequency data}
This component takes time-frequency $\mathcal{F}$ as an input. $\mathcal{F}$ is generated by applying Short Time Fourier Transform (STFT) to the time-series $\mathcal{X}$. The equation for STFT is shown in Eq:\ref{eq:STFT} where $\mathcal{S}$ is the STFT function and $g(n-m)$ is the window function. Usually a Hann or a Gaussian window is used and the width of window is specified by $m$. 
\vspace{-2mm}
\begin{equation}
\begin{split}
\{s_i\}_{i=1}^{n} & = \mathcal{S}(\{\mathcal{X}_i\}_{i=1}^{n}) \\
\mathcal{S}(\{\mathcal{X}_i\})(k,m)  & = \sum_{n=0}^{N-1} \mathcal{X}(n)g(n-m)e^{\frac{-j2\pi k n}{N}} 
\end{split}
\label{eq:STFT}
\end{equation}

 where $k$ and $m$ are the time index and frequency index, respectively. The time-frequency responses $\{s_i\}_{i=1}^{i=n}$ is passed through transformations $\mathbb{T}^2$, (i.e. $\mathbb{T}^2_1$ and $\mathbb{T}^2_2$ are global and local transformations). The SSL KD in this task is performed by Time Cutout (TC) and Frequency Cutout (FC) transformation $\mathbb{T}^{2}$ as follows:
 \vspace{-2mm}
 \begin{equation}
     \mathbb{T}^{2} : \{F_{x,y}\}_{x={t}_{1},y={f}_{1}}^{{t}_{2},{f}_{2}} = 0
 \end{equation}
 where $\{F_{x,y}\}_{x={t}_{1}}^{{t}_{2}} = 0$ and $\{F_{x,y}\}_{y={f}_{1}}^{{f}_{2}} = 0$ are TC and FC transformations, respectively.  
 

 
We utilize SE-ResNet34, a modified version of ResNet \cite{he2016deep} as our backbone network for both $\mathcal{T}^2_{\hat{\phi}}$ and $\mathcal{S}^2_{\hat{\theta}}$ which are trained on different parameters of $\hat{\phi}$ and $\hat{\theta}$.

\subsection{SSL: Down-stream Task}
Different from pre-stream task trained on unlabeled data, the down-stream task is trained on labeled data. Given a signal $\mathcal{X}$, we transfer student models (i.e., $\mathcal{S}^1_\theta$, $\mathcal{S}^2_{\hat{\theta}}$), which were unsupervised trained by the previous module, to extract ECG characteristic from time-series ($f_1 = \mathcal{S}^1_{\theta}(\mathcal{X})$) and time-frequency signal ($f_2 = \mathcal{S}^2_{\hat{\theta}}(\mathcal{X})$). Because features $f_1$ and $f_2$ carry different characteristic of $\mathcal{X}$, we propose a \textbf{gate fusion mechanism} as follows: (i) We first process features $f_1$ and $f_2$ through a soft attention \cite{xu2015show} to learn impact factors ($w_1$, $w_2$) of each feature. The output are weighted and denoted as $\hat{f}_1$ and $\hat{f}_2$. (ii) We then stack features $\hat{f}_1$ and $\hat{f}_2$ and form $f = [\hat{f}_1$, $\hat{f}_2]$. A multi-layer perceptron (MLP) and sigmoid are then applied into $f$. (iii) We finally train the down-stream task with a binary cross entropy loss between the predicted label output from the sigmoid function and groundtruth label.

\section{Experimental Results}

\noindent
\textbf{A. Dataset, Metrics}
We conduct our experiments on PhysioNet 2020 dataset \cite{alday2020classification}. PhysioNet comprises six ECG datasets and the total number of records are 43,101 for 111 diagnoses. To fairly compare with other existing methods, we adopt data stratification from \cite{mehari2022self} and only 25 classes are selected according to the challenge. Our proposed network is trained using  ten-fold cross validation. For a broader analysis of performance 
the results are described under all recommended metrics i.e., AUROC, AUPRC, Accuracy (Acc), F1, F2, G-score and the challenge metric (ChM) as in \cite{alday2020classification}.

\begin{table}[htbp]	
\caption{Performance comparison between proposed method and other approaches on the PhysioNet2020. SSL-T, SSL-S and  SSL-TSG are different setups of our network.}
\resizebox{\linewidth}{!}{
\begin{tabular}{c|c|c|c|c|c|c|c|c}
\hline
& \textbf{\textit{Model}} & \textbf{\textit{AUROC}} $\uparrow$& \textbf{\textit{AUPRC}}$\uparrow$& \textbf{\textit{Acc}} $\uparrow$& \textbf{\textit{F1}} $\uparrow$& \textbf{\textit{F2}} $\uparrow$& \textbf{\textit{G2}} $\uparrow$& \textbf{\textit{ChM}}$\uparrow$ \\
\hline
\multirow{9}{*}{\rotatebox{90}{Existing Works}} &\cite{ignacio2020topology} &  84.6 & 32.2 & 20.9  & 25.5 & -- & -- & 21.9  \\
& \cite{duan2020madnn} & -- & -- & -- & -- & -- & -- & 23.6  \\
& \cite{wong2021multilabel} & 80.6 & 26.1 & 11.3 & 26.0  & 30.9  & 12.6  & 24.8 \\
& \cite{nankani2020automatic} & 82.5 & 32.6 & 33.1 & 28.6  & -- & -- & 30.5  \\
& \cite{singstad2020convolutional} & 87.2 & -- & --  & 39.9  & 43.6  & 23.7  & 40.9  \\ 
& \cite{feng2020deep}  & 55.1 & 52.3 & -- & 53.8 & -- & -- & 53.9 \\
& \cite{natarajan2020wide}  & --  & -- & -- & -- & -- & -- & 53.3 \\
& \cite{nonaka2020electrocardiogram} & --  & -- & -- & -- & -- & -- & 58.5 \\
& \cite{jiang2020diagnostic} & -- & -- & --  & 60.3 & -- & -- & 62.7 \\
\hline
\multirow{3}{*}{\rotatebox{90}{Our}}& SSL-T & 95.7 & 64.8 & 48.8 & 61.4 & 64.9 & 40.2 & 64.7 \\
& SSL-S & 93.8 & 56.3 & 40.7 & 53.2 & 57.4 & 33.9 & 58.0 \\
& SSL-TSG & \textbf{96.0} & \textbf{65.8} & \textbf{48.9} & \textbf{62.1} & \textbf{65.9} & \textbf{41.0} & \textbf{65.4} \\
\hline
\end{tabular}}
\label{tab1}
\end{table}

\noindent
\textbf{B. Performance and Comparison}
We conduct a comparison between our proposed method with other existing approaches as shown in Table~\ref{tab1}. In this experiment, we report the performance of our proposed SSL with three different setups: (i) SSL-T: Work on one modality of time-series data. The network is $\mathcal{S}^1_\theta$. There is no gate fusion used; (ii) SSL-S: Work on one modality of time-frequency data. The network is $\mathcal{S}^2_{\hat{\theta}}$. There is no gate fusion used; (iii) SSL-TSG: Our proposed method. Both modalities are used. Gate fusion is needed to fuse $f_1$ and $f_2$. Our proposed method obtains the best performance and it gains
significant margins on various metrics compared with the existing methods.

The ablation study in Table~\ref{tab2} contains two groups corresponding to the utilization of SSL or not. In the case of w/o SSL, the networks $\mathcal{S}^1_\theta$ and $\mathcal{S}^2_{\hat{\theta}}$ are not pre-trained by self-KD. At each group, we test the network on various settings: (i) T: One modality of time-series data with network $\mathcal{S}^1_\theta$. (ii) S: One modality of time-frequency data with network $\mathcal{S}^2_{\hat{\theta}}$. (iii) TSC: Both modalities with both networks $\mathcal{S}^1_\theta$ and $\mathcal{S}^2_{\hat{\theta}}$ and there is no gate fusion used. The feature $f_1$ and $f_2$ are combined by concatenation; (iv) TSG: Both modalities with both networks $\mathcal{S}^1_\theta$ and $\mathcal{S}^2_{\hat{\theta}}$. The feature $f_1$ and $f_2$ are combined by our proposed gate fusion mechanism. On each group of with or without SSL, even time-series outperforms time-frequency, the combination of both modalities with our gate fusion yields the best performance.
The effectiveness of SSL (i.e self-KD) can be clearly seen when comparing the results with or without SSL on the same modality.



\begin{table}[htbp]
\caption{Ablation study of proposed method with different settings. In the $2^{nd}$ group, SSL denotes when Self-KD is used.}
\resizebox{\linewidth}{!}{
\begin{tabular}{|c|c|c|c|c|c|c|c|}
\hline
\textbf{} & \textbf{\textit{AUROC}}& \textbf{\textit{AUPRC}}& \textbf{\textit{Acc}} & \textbf{\textit{F1}} & \textbf{\textit{F2}} & \textbf{\textit{G2}} & \textbf{\textit{ChM}} \\
\hline
T & 95.34 & 62.93 & 46.14 & 59.56 & 63.37 & 38.43 & 63.25 \\ 
S & 93.47 & 54.98 & 39.53 & 52.17 & 56.31 & 32.85 & 56.78 \\
TSC & 95.41 & 63.77 & 47.13 & 60.54 & 64.12 & 39.49 & 63.72 \\
TSG & 95.57 & 64.22 & 48.04 & 61.01 & 64.18 & 39.74 & 64.16 \\
\hline
SSL-T & 95.70 & 64.81 & 48.78 & 61.40 & 64.89 & 40.17 & 64.71 \\
SSL-S & 93.76 & 56.29 & 40.68 & 53.22 & 57.44 & 33.93 & 58.01 \\
SSL-TSC & 95.65 & 64.64 & 48.13 & 60.92 & 64.33 & 39.73 & 64.37 \\
SSL-TSG & \textbf{95.99} & \textbf{65.84} & \textbf{48.91} & \textbf{62.07} & \textbf{65.95} & \textbf{40.98} & \textbf{65.39} \\
\hline
\end{tabular}}
\label{tab2}
\end{table}

\begin{figure}[!h]	
\centering
\includegraphics[width=0.45\textwidth]{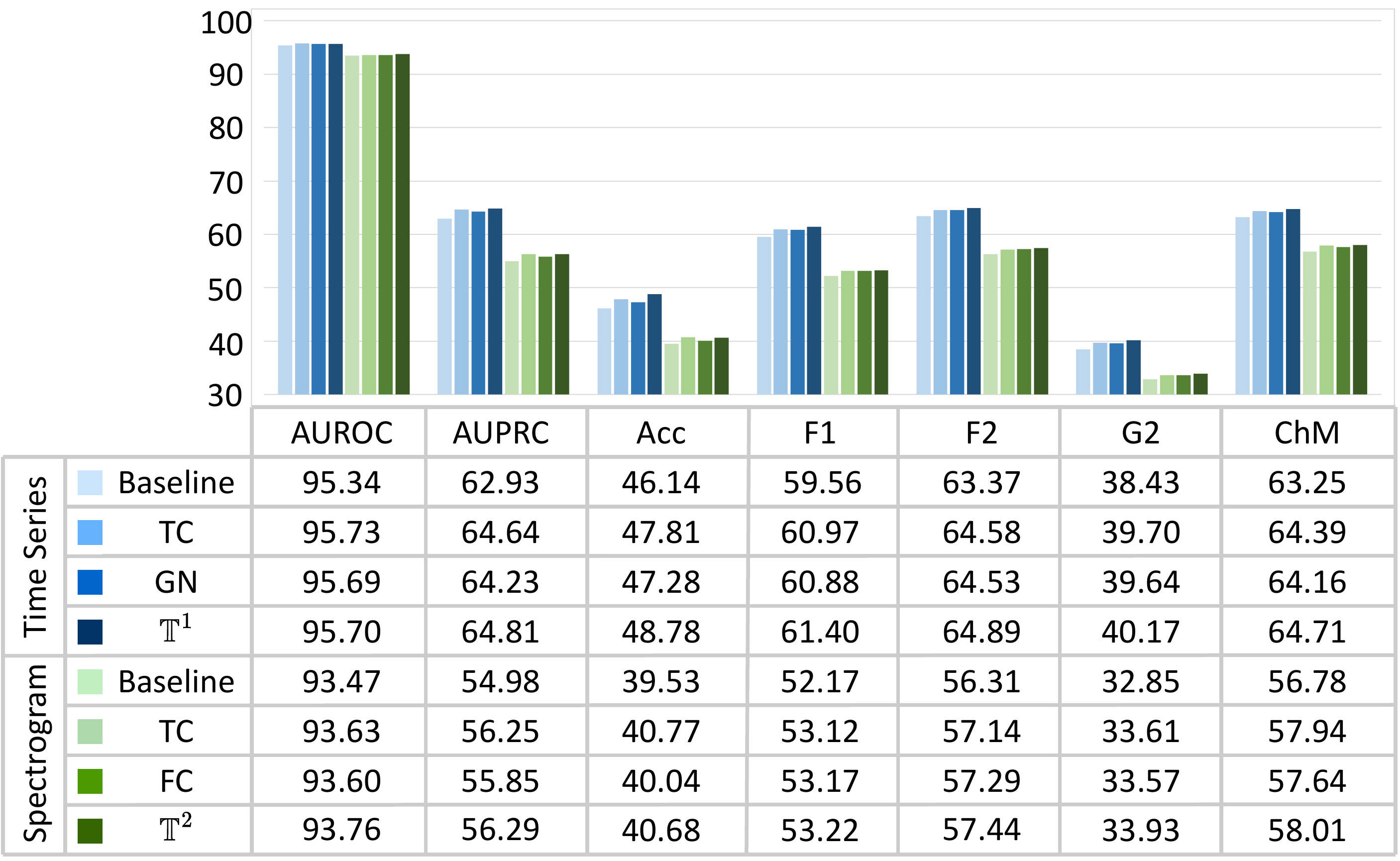}
\vspace{-4mm}
\caption{Ablation study on different transformation setup on each modality. \vspace{-4mm}}
\label{fig:table3}
\end{figure}

Transformations are crucial to the success of the adopted SSL \cite{caron2021emerging}; hence, we further conduct an ablation study on different transformations as shown in Fig.~\ref{fig:table3}. On each domain, a baseline is defined when no transformation is used. 
In time series domain, we report the performance on transformation of TC, GN and $\mathbb{T}^1$ as defined in Subsec.\ref{subsec:pre}.1. In time-frequency domain, we report the performance on transformation of TC, FC and $\mathbb{T}^2$ as defined in Subsec.\ref{subsec:pre}.2. The baseline obtains good performance -- thanks to the inherent temporal dependency modeling in our network design. Given a long ECG record, most of existing work pre-processed the ECG signal by  partitioning it into a set of overlapping segments. In such approaches, the temporal relationship between segments mainly depends on the overlap threshold, which is typically set as a pre-defined parameter. Unlike the existing work, we randomly crop the ECG record into a set of segments at each iteration, thus, the overlap between segments is more flexible and the inter-segment coherence is modeled better.

\vspace{-4mm}
\section*{Conclusion}
In this paper, we have proposed an SSL-based multimodality network for multi-lead ECG classification. Our approach includes two modules, one for the pre-stream task to pre-train networks on unlabeled data, and the other on the down-stream task to fine-tune the networks on labeled data. Our SSL is built based on a self-KD mechanism. We have investigated two modalities corresponding respectively to time-series data, and spectrogram in time-frequency domain. To combine features from multiple modalities, we have introduced a gate fusion mechanism. The experimental results and an extensive ablation study have shown that SSL on multi-modality data is an effective approach to classify multi-lead ECG. 

\noindent
\textbf{Acknowledgments:}
This material is based upon work supported in part by the US National Science Foundation, under Award No. OIA-1946391, NSF 1920920.


{\footnotesize
\bibliographystyle{unsrt} 
\bibliography{refs}

\begin{thebibliography}{10}

\bibitem{chouhan2008threshold}
V~Chouhan and S~Mehta.
\newblock Threshold-based detection of p and t-wave in ecg using new feature
  signal.
\newblock {\em Com.Sci.\& Net.Sec.}, 8(2):144--153, 2008.

\bibitem{zabihi2017detection}
M~Zabihi, A~Rad, et~al.
\newblock Detection of atrial fibrillation in ecg hand-held devices using a
  random forest classifier.
\newblock In {\em CinC}, pages 1--4, 2017.

\bibitem{yamazaki2022spiking}
Kashu Yamazaki, Viet-Khoa Vo-Ho, Darshan Bulsara, and Ngan Le.
\newblock Spiking neural networks and their applications: A review.
\newblock {\em Brain Sciences}, 12(7):863, 2022.

\bibitem{le2021deep}
Ngan Le, Vidhiwar~Singh Rathour, Kashu Yamazaki, Khoa Luu, and Marios Savvides.
\newblock Deep reinforcement learning in computer vision: a comprehensive
  survey.
\newblock {\em Artificial Intelligence Review}, pages 1--87, 2021.

\bibitem{zhou2021deep}
S~Kevin Zhou, Hoang~Ngan Le, Khoa Luu, Hien~V Nguyen, and Nicholas Ayache.
\newblock Deep reinforcement learning in medical imaging: A literature review.
\newblock {\em Medical image analysis}, 73:102193, 2021.

\bibitem{rajpurkar2017cardiologist}
P~Rajpurkar, A.Y Hannun, et~al.
\newblock Cardiologist-level arrhythmia detection with convolutional neural
  networks.
\newblock {\em arXiv preprint arXiv:1707.01836}, 2017.

\bibitem{le2021multi}
M.D Le, V.S Rathour, et~al.
\newblock Multi-module recurrent convolutional neural net with transformer
  encoder for ecg arrhythmia classification.
\newblock In {\em BHI}, pages 1--5. IEEE, 2021.

\bibitem{ribeiro2020automatic}
A.H Ribeiro, M.Horta Ribeiro, G.M.M Paix{\~a}o, D.M Oliveira, P.R Gomes, J.A
  Canazart, M.P.S Ferreira, C.R Andersson, P.W Macfarlane, W~Meira~Jr, et~al.
\newblock Automatic diagnosis of the 12-lead ecg using a deep neural network.
\newblock {\em Nature communications}, 11(1):1--9, 2020.

\bibitem{baloglu2019classification}
U.B Baloglu, M~Talo, O~Yildirim, R~San~Tan, and U.R Acharya.
\newblock Classification of myocardial infarction with multi-lead ecg signals
  and deep cnn.
\newblock {\em Pattern Recognition Letters}, 122:23--30, 2019.

\bibitem{zhang2016colorful}
R~Zhang, P~Isola, and A.A Efros.
\newblock Colorful image colorization.
\newblock In {\em ECCV}, pages 649--666. Springer, 2016.

\bibitem{wang2019global}
G~Wang, C~Zhang, Y~Liu, H~Yang, D~Fu, H~Wang, and P~Zhang.
\newblock A global and updatable ecg beat classification system based on
  recurrent neural networks and active learning.
\newblock {\em Information Sciences}, 501:523--542, 2019.

\bibitem{huang2019ecg}
J~Huang, B~Chen, B~Yao, and W~He.
\newblock Ecg arrhythmia classification using stft-based spectrogram and
  convolutional neural network.
\newblock {\em IEEE access}, 7:92871--92880, 2019.

\bibitem{yildirim2018novel}
{\"O}~Yildirim.
\newblock A novel wavelet sequence based on deep bidirectional lstm net model
  for ecg signal classification.
\newblock {\em CBM}, 96:189--202, 2018.

\bibitem{caron2021emerging}
M~Caron, H~Touvron, I~Misra, H~J{\'e}gou, J~Mairal, P~Bojanowski, and A~Joulin.
\newblock Emerging properties in self-supervised vision transformers.
\newblock In {\em Proceedings of the IEEE/CVF ICCV}, pages 9650--9660, 2021.

\bibitem{caron2018deep}
M~Caron, P~Bojanowski, A~Joulin, and M~Douze.
\newblock Deep clustering for unsupervised learning of visual features.
\newblock In {\em ECCV}, pages 132--149, 2018.

\bibitem{srivastava2015unsupervised}
N~Srivastava, E~Mansimov, and R~Salakhudinov.
\newblock Unsupervised learning of video representations using lstms.
\newblock In {\em ICML}, pages 843--852. PMLR, 2015.

\bibitem{pathak2017learning}
D~Pathak, R~Girshick, P~Doll{\'a}r, T~Darrell, and B~Hariharan.
\newblock Learning features by watching objects move.
\newblock In {\em CVPR}, pages 2701--2710, 2017.

\bibitem{tran2022ss}
M~Tran, L~Ly, B.S Hua, and N~Le.
\newblock Ss-3dcapsnet: Self-supervised 3d capsule networks for medical
  segmentation on less labeled data.
\newblock {\em IEEE ISBI}, 2022.

\bibitem{sayed2018cross}
N~Sayed, B~Brattoli, and B~Ommer.
\newblock Cross and learn: Cross-modal self-supervision.
\newblock In {\em GCPR}, pages 228--243, 2018.

\bibitem{vu2021self}
D.Q Vu, N.T.H Le, and J.C Wang.
\newblock Self-supervised learning via multi-transformation classification for
  action recognition.
\newblock {\em arXiv preprint arXiv:2102.10378}, 2021.

\bibitem{kiyasseh2020clocs}
D~Kiyasseh, T~Zhu, and D.A Clifton.
\newblock Clocs: Contrastive learning of cardiac signals.
\newblock {\em arXiv preprint arXiv:2005.13249}, 2020.

\bibitem{lan2021intra}
X~Lan, D~Ng, S~Hong, and M~Feng.
\newblock Intra-inter subject self-supervised learning for multivariate cardiac
  signals.
\newblock {\em arXiv preprint arXiv:2109.08908}, 2021.

\bibitem{mehari2022self}
T~Mehari and N~Strodthoff.
\newblock Self-supervised representation learning from 12-lead ecg data.
\newblock {\em CBM}, 141:105114, 2022.

\bibitem{vu2021teaching}
Duc-Quang Vu, Ngan Le, and Jia-Ching Wang.
\newblock Teaching yourself: A self-knowledge distillation approach to action
  recognition.
\newblock {\em IEEE Access}, 9:105711--105723, 2021.

\bibitem{hinton2015distilling}
G~Hinton, O~Vinyals, J~Dean, et~al.
\newblock Distilling the knowledge in a neural network.
\newblock {\em arXiv preprint arXiv:1503.02531}, 2(7), 2015.

\bibitem{he2019bag}
T~He, Z~Zhang, H~Zhang, X~Zhang, J~Xie, and M~Li.
\newblock Bag of tricks for image classification with convolutional neural
  networks.
\newblock In {\em CVPR}, pages 558--567, 2019.

\bibitem{he2016deep}
Kaiming He, Xiangyu Zhang, Shaoqing Ren, and Jian Sun.
\newblock Deep residual learning for image recognition.
\newblock In {\em CVPR}, pages 770--778, 2016.

\bibitem{xu2015show}
K~Xu, J~Ba, R~Kiros, K~Cho, A~Courville, R~Salakhudinov, R~Zemel, and Y~Bengio.
\newblock Show, attend and tell: Neural image caption generation with visual
  attention.
\newblock In {\em ICML}, pages 2048--2057. PMLR, 2015.

\bibitem{alday2020classification}
E.A.P Alday, A~Gu, A.J Shah, C~Robichaux, A.K.I Wong, C~Liu, F~Liu, A.B Rad,
  A~Elola, S~Seyedi, et~al.
\newblock Classification of 12-lead ecgs: the physionet/computing in cardiology
  challenge 2020.
\newblock {\em Physiological measurement}, 41(12):124003, 2020.

\bibitem{ignacio2020topology}
P.S Ignacio, J.A Bulauan, and J.R Manzanares.
\newblock A topology informed random forest classifier for ecg classification.
\newblock In {\em 2020 CinC}, pages 1--4, 2020.

\bibitem{duan2020madnn}
R~Duan, X~He, and Z~Ouyang.
\newblock Madnn: A multi-scale attention deep neural net for arrhythmia
  classification.
\newblock In {\em 2020 CinC}, pages 1--4, 2020.

\bibitem{wong2021multilabel}
A.W Wong, A~Salimi, et~al.
\newblock Multilabel 12-lead electrocardiogram classification using beat to
  sequence autoencoders.
\newblock In {\em ICASSP}, pages 1270--1274. IEEE, 2021.

\bibitem{nankani2020automatic}
D~Nankani, P~Saikia, and R.D Baruah.
\newblock Automatic concurrent arrhythmia classification using deep residual
  neural networks.
\newblock In {\em 2020 CinC}, pages 1--4, 2020.

\bibitem{singstad2020convolutional}
B.J Singstad and C~Tronstad.
\newblock Convolutional neural net and rule-based algorithms for classifying
  12-lead ecgs.
\newblock In {\em 2020 CinC}, pages 1--4, 2020.

\bibitem{feng2020deep}
Y~Feng and E~Vigmond.
\newblock Deep multi-label multi-instance classification on 12-lead ecg.
\newblock In {\em 2020 CinC}, pages 1--4, 2020.

\bibitem{natarajan2020wide}
A~Natarajan, Y~Chang, et~al.
\newblock A wide and deep transformer neural network for 12-lead ecg
  classification.
\newblock In {\em 2020 CinC}, pages 1--4, 2020.

\bibitem{nonaka2020electrocardiogram}
N~Nonaka and J~Seita.
\newblock Electrocardiogram classification by modified efficientnet with data
  augmentation.
\newblock In {\em 2020 CinC}, pages 1--4, 2020.

\bibitem{jiang2020diagnostic}
Z~Jiang, T.P Almeida, et~al.
\newblock Diagnostic of multiple cardiac disorders from 12-lead ecgs using
  graph convolutional net based multi-label classification.
\newblock In {\em 2020 CinC}, pages 1--4, 2020.

\end{thebibliography}
}

\end{document}